\documentclass[sigconf, natbib=true]{acmart}
\usepackage[normalem]{ulem}
\useunder{\uline}{\ul}{}
\usepackage{graphicx}
\usepackage{subcaption}
\usepackage{amsmath}
\usepackage{enumitem}
\usepackage{xspace}
\theoremstyle{definition}

\newcommand{\modelname}{PRTA\xspace}
\AtBeginDocument{%
  }

\copyrightyear{2026}
\acmYear{2026}
\setcopyright{cc}
\setcctype{by}
\acmConference[RecSys '26]{20th ACM Conference on Recommender Systems}{September 27-October 02, 2026}{Minneapolis, MN, USA}
\acmBooktitle{20th ACM Conference on Recommender Systems (RecSys '26), September 27-October 02, 2026, Minneapolis, MN, USA}
\acmDOI{10.1145/3773078.3831794}
\acmISBN{979-8-4007-2284-4/2026/09}

\begin{document}

%%
%% The "title" command has an optional parameter,
%% allowing the author to define a "short title" to be used in page headers.
% \title{Cross-view Graph Self-supervised Learning for Group Identification via Transitional Hypergraph Convolution}
% \title{Unified Multitask pretraining for Recommendation via Hypergraph with Transitional Attention}
\title{Personalized Recommendation Tool Learning via Autonomous Language Agents}
\titlenote{This short paper focuses on the personalized tool-learning component of our broader study; the extended version, which additionally infers user intent over substitute and complement items via LLM reasoning, appears in AgentDR~\cite{agentdr}.}

\author{Mingdai~Yang}
\email{myang72@uic.edu}
\affiliation{%
  \institution{Univ.\ of Illinois  Chicago}
  \city{Chicago}
  \state{IL}
  \country{USA}}

\author{Zhiwei~Liu}
\email{zhiweiliu@microsoft.com}
\affiliation{%
  \institution{Microsoft}
  \city{Mountain View}
  \country{USA}
}

\author{Weizhi~Zhang}
\email{wzhan42@uic.edu}
\affiliation{%
  \institution{Univ.\ of Illinois  Chicago}
  \city{Chicago}
  \state{IL}
  \country{USA}}

\author{Yibo~Wang}
\email{ywang633@uic.edu}
\affiliation{%
  \institution{Univ.\ of Illinois  Chicago}
  \city{Chicago}
  \state{IL}
  \country{USA}}

\author{Hao Peng}
\orcid{0000-0003-0458-5977}
\email{penghao@buaa.edu.cn}
\affiliation{%
   \institution{Beihang University,~\& Hangzhou Innovation Institute of BUAA}
   \country{Beijing~\& Hangzhou, China}}
\authornote{Corresponding author}

\author{Philip Yu}
\email{psyu@uic.edu}
\affiliation{%
  \institution{Univ.\ of Illinois  Chicago}
  \city{Chicago}
  \state{IL}
  \country{USA}}  

%%
%% The abstract is a short summary of the work to be presented in the
%% article.
\begin{abstract}
Although large language models (LLMs) have recently gained traction in recommender systems due to their strong reasoning capabilities and extensive world knowledge, previous LLM-based agents suffer from hallucination and context-length limitations, and thus are not suitable for full-ranking recommendation tasks. To circumvent these limitations through architectural design rather than modifying the LLM itself, we propose an agent-based recommendation framework, memory-based $\textbf{P}$ersonalized $\textbf{R}$ecommendation $\textbf{T}$ool learning via autonomous language $\textbf{A}$gents (PRTA), in which an LLM acts as a central planner interacting with multiple recommendation models as tools. The LLM-based agent is responsible for high-level reasoning and personalized tool selection, while traditional recommendation models perform full-ranking scoring, leveraging their scalability in modeling behavioral patterns. To support personalized tool selection, we design reflection mechanisms that enable the agent to evaluate and compare tools for each user based on user profiles and candidate ranked lists. Extensive experiments across three public datasets demonstrate the superiority of \modelname over traditional recommendation and LLM-based baselines in improving full-ranking recommendation performance. Our code implementation is available online\footnote{https://github.com/mdyfrank/RecToolAgent}. 
\end{abstract}

%%
%% The code below is generated by the tool at http://dl.acm.org/ccs.cfm.
%% Please copy and paste the code instead of the example below.
%%


%%
%% Keywords. The author(s) should pick words that accurately describe
%% the work being presented. Separate the keywords with commas.

%% A "teaser" image appears between the author and affiliation
%% information and the body of the document, and typically spans the
%% page.

%%
%% This command processes the author and affiliation and title
%% information and builds the first part of the formatted document.
\maketitle

\section{Introduction}
Traditional embedding-based recommendation models primarily rely on historical user–item interaction data. Even when incorporating LLMs to encode textual features into embeddings~\cite{ihp,pcl}, such representations remain limited in capturing nuanced, context-dependent semantics and rich world knowledge~\cite{WuZHHYCISLZ24,0007XCW25}.
To address this limitation, prior works have explored using LLMs directly as recommender systems by formulating recommendation tasks as language processing problems~\cite{p5,tallrec,HuaGXJLZ24}. However, the inherent gap between language modeling and personalized behavior modeling prevents standalone LLMs from effectively capturing collaborative signals embedded in user–item interactions~\cite{agentcf}. Moreover, such approaches typically require substantial computational resources for fine-tuning LLMs.
To overcome these drawbacks, a rapidly growing body of work explores using LLMs as interactive agents for recommendation~\cite{recmind,interecagent,agent4rec,recagent,iagent}. Beyond prompting LLMs with user interaction histories, these inference-only methods incorporate memory modules to preserve user context and better exploit collaborative signals.

However, additional efforts are needed to alleviate hallucination and token limitation in LLMs~\cite{JiangWMCWWZ25,JiLFYSXIBMF23,agentdr}.
Specifically, LLMs may generate non-existent items or produce item descriptions that cannot be reliably mapped to valid item IDs, unlike traditional recommender systems that rank items from a fixed catalog.
There have been attempts to constrain the text output space using carefully crafted query prompts~\cite{agentcf, agent4rec}, but their effectiveness depends heavily on the pretrained LLM’s ability to strictly follow instructions. Meanwhile, the inherent token length limitation of LLMs prevents them from ranking over the entire item catalog.  As a result, prior agent-based works evaluate their performance by ranking the target item against a small set of sampled negative items, deviating from practical recommendation demands~\cite{agentcf,recmind,interecagent,agent4rec,recagent,iagent}. 

Rather than attempting to resolve hallucination and context-length constraints \emph{within} the LLM itself, e.g., via fine-tuning or elaborate instruction-following prompts, we adopt an architectural engineering perspective that \emph{circumvents} these limitations by design. Specifically, instead of relying on the instruction-following ability of LLMs or invoking external validation tools, we argue that naturally valid outputs can be achieved by orchestrating multiple recommendation models as tools within an agent framework. Inspired by recent advances in tool learning for LLMs~\cite{Toolformer,HuggingGPT,ReAct,WuZHHYCISLZ24}, and to combine the strengths of traditional recommendation models with LLM-based reasoning, we propose an agent-based recommendation framework \modelname. In this framework, behavioral patterns are primarily captured by diverse recommendation models, while the LLM-based agent focuses on reasoning about how to selectively utilize these tools based on its language understanding of textual information about users and items. Unlike existing agent-based frameworks~\cite{toolrec,agentcf}, \modelname leverages multiple recommendation paradigms rather than relying on a single one, thereby enabling more robust and comprehensive recommendation.
The key contributions of this work are as follows:
\begin{itemize}[left=0pt]
    \item We propose \modelname, a novel agent-based recommendation framework that decouples behavioral modeling from semantic reasoning by delegating full-ranking inference to specialized recommendation models while using an LLM agent for tool selection.
    \item We design three reflection mechanisms and a lightweight reranking module to integrate tools and refine top-ranked items.
    \item Extensive experiments on three public datasets demonstrate the effectiveness of \modelname under full-ranking evaluation.
\end{itemize}

\section{Preliminaries}
\subsection{Problem formulation} Let $\mathcal{I}$ denote the complete item set and $\mathcal{U}$ denote the set of users. Each user $u \in \mathcal{U}$ is associated with a behavior sequence $\mathbf{s} = (i_1, i_2, \dots, i_n)$, where $i_j \in \mathcal{I}$ denotes the $j$-th item interacted with by $u$. Each item is also accompanied by a textual description. For simplicity, we use $\text{desc}(\mathbf{s})$ to denote the list of textual descriptions of items in $\mathbf{s}$. The goal is to predict the next item $i_{n+1} \in \mathcal{I}$ that the user will interact with, under a full-ranking setting where all items in $\mathcal{I}$ are ranked for each user.

\subsection{Recommendation tools} \modelname is compatible with any recommendation model that outputs a full-ranking list of item scores for each user, to mitigate hallucination from LLM outputs. In this work, we adopt three pretrained recommenders as tools: LightGCN~\cite{LightGCN} with score list $\hat{\mathbf{r}}_{G}$, SASRec~\cite{sasrec} with score list $\hat{\mathbf{r}}_{S}$, and SimpleX~\cite{simplex} with score list $\hat{\mathbf{r}}_{M}$. The tool set $\mathcal{T}$, including these three tools, is representative for a clean evaluation of our framework without relying on complex recommendation tools.

\section{Proposed Framework: PRTA}
\begin{figure*}[]
    \centering
    \includegraphics[scale=0.55]{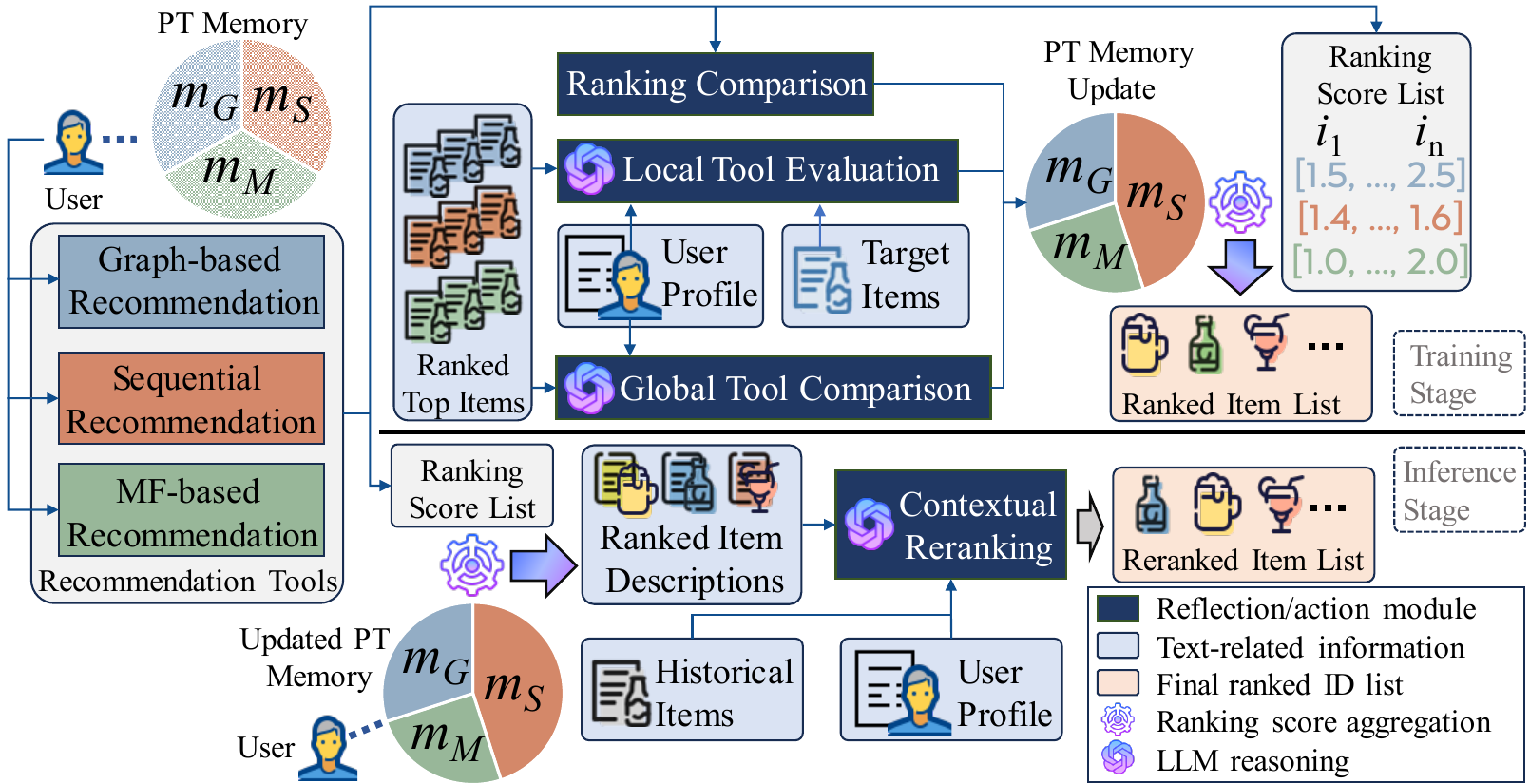}
    \caption{The framework of \modelname. Each user agent is equipped with a PT memory that stores personalized tool preferences, optimized through local tool evaluation, global tool comparison, and ranking comparison. 
    During inference, ranking scores from multiple tools are aggregated based on the PT memory, and the top items are refined by a contextual reranking module.
    }
    \label{fig:framework}
\end{figure*}
\subsection{User Profile and Agent Memory}
\subsubsection{User Profile Generation}
First, we prompt the LLM to summarize the user's profile based on descriptions of historical items:
\begin{equation}
        y_{\text{prof}} = \text{LLM}_\text{prof}(\text{desc}(\mathbf{s}_{[:-k]}))
\end{equation}
where desc($\mathbf{s}_{[:-k]}$) denotes the sequence of textual descriptions associated with the user’s pretrained historical interaction sequence $\mathbf{s}$, and $\text{LLM}_\text{prof}(\cdot)$ represents the LLM prompted to transform these texts into a coherent user profile. As shown in Fig.~\ref{fig:framework}, the user profile is used for contextual reranking in Sec.~\ref{sec:rerank}. 

\subsubsection{Personalized Tool Memory} To enable interpretable tool selection among diverse recommendation models, we equip the agent with a compact Personalized Tool (PT) memory. Specifically, PT memory $\mathbf{m}\in\mathbb{R}^{1\times d}$ is a weight vector that describes the suitability of each tool for each user, where $d$ is the number of tools deployed in the framework. For LightGCN, SASRec, and SimpleX deployed as tools, we denote their weights as $m_G$, $m_S$, and $m_M$, respectively.
In ranking aggregation, a weighted sum is computed to generate the final ranking list $\hat{\mathbf{r}}\in\mathbb{R}^{1\times|\mathcal{I}|}$ for each user based on the PT memory:
\begin{equation}
    \hat{\mathbf{r}} = \sum_{t\in\mathcal{T}}m_t\hat{\mathbf{r}}_t = m_G\hat{\mathbf{r}}_{G} +m_S\hat{\mathbf{r}}_{S} + m_M\hat{\mathbf{r}}_{M}.
\end{equation}
This weighted aggregation allows the agent to personalize the contribution of each tool according to the learned PT memory. It not only enhances recommendation accuracy but also maintains interpretability, as the tool weights offer insight into the agent's decision-making process.

\subsection{Agent Reflection}\label{sec:reflection}
\subsubsection{Local Tool Evaluation}
Given the user agents with profiles and memories, our goal is to optimize them to simulate real-world users’ selection of tools. To achieve this, we judge the suitability of each tool based on relevance between the top items in its ranked list for each user and the textual descriptions of items in the user's interaction history. We use $f_\text{top}(\hat{\mathbf{r}}_{t}, k_{\text{eval}})$ to represent the list of top $k_{\text{eval}}$ items ranked by a tool for user $u$. Then the agent is tasked with determining the relevance between this list and the items in the behavior sequence of this user:
\begin{equation}
    y_{\text{eval}}^{t} = \text{LLM}_{\text{eval}}( y_{\text{prof}}, \text{desc}(f_\text{top}(\hat{\mathbf{r}}_{t}, k_{\text{eval}})), \text{desc}(\mathbf{s}_{[-k:]}))
\end{equation}
where $\mathbf{s}_{[-k:]}$ denotes the most recent $k$ items interacted with by the user that are not visible to the tools during pretraining, and $y_{\text{eval}}^{t}\in\{-1,0,1\}$ denotes the relevance label, where $-1$ indicates irrelevant, $0$ indicates neutral, and $1$ indicates relevant. In our implementation, we prompt the LLM to generate relevance as text output, which is then mapped to numerical values, as language models tend to reason and express information more accurately in natural language. This process is represented as the function $\text{LLM}_{\text{eval}}(\cdot)$.
Based on the relevance between each tool’s ranked list and the user’s behavior sequence, we update the corresponding tool weight $m_t \in \{m_G, m_S, m_M\}$ in the PT memory:
\begin{equation}
    m_{t} \leftarrow  m_{t} + \alpha\cdot y_{\text{eval}}^{t} 
\end{equation}
where $\alpha$ is the learning rate to update PT memory by this local tool evaluation. This local tool evaluation process enables each agent to iteratively refine its PT memory through grounding tool suitability in behavioral relevance.

\subsubsection{Global Tool Comparison}
Local tool evaluations are performed independently without explicitly considering trade-offs between tools. To enable a more holistic and personalized comparison that captures the complementary strengths of different models, we further propose a global tool comparison module. In this module, the agent is tasked with selecting the most relevant tool from $\mathcal{T}$:
\begin{equation}
     y_{\text{comp}}^{t} = \text{LLM}_{\text{comp}}(\{\text{desc}(f_\text{top}(\hat{\mathbf{r}}_{t}, k_{\text{comp}})) \mid t\in \mathcal{T}\}, \text{desc}(\mathbf{s}_{[-k:]}))
\end{equation}
where $y_{\text{comp}}^{t}\in \{0,1\}$ is a binary indicator representing whether tool $t$ produces the ranked list most relevant to the user’s recent behavior sequence among all candidates. In our implementation, we prompt the LLM to select the best tool in natural language and then convert its output into the corresponding binary indicators, which is denoted as $\text{LLM}_{\text{comp}}(\cdot)$.
Similar to local tool evaluation, the PT memory is updated based on $y_{\text{comp}}^{t}$:
\begin{equation}
    m_{t} \leftarrow  m_{t} + \beta\cdot y_{\text{comp}}^{t} 
\end{equation}
where $\beta$ is the learning rate. The global tool comparison provides the agent with a broader preference-aligned perspective across tools, going beyond isolated evaluations.

\subsubsection{Ranking Comparison} 
In addition to the LLM-based global tool comparison, we compare ranking performance across tools based on the ranks of the most recent user-interacted items that were not visible to the tools during ranking. This complementary signal mitigates potential bias in the LLM's textual judgments and improves the reliability of tool selection. The PT memory is updated based on the ranks $\{r^j_{t}\mid i_j\in\mathbf{s}_{[-k:]}\}$ of the most recent items:
\begin{equation}
    m_{t} \leftarrow  m_{t} + \sum_{\{r^j_{t}\mid i_j\in\mathbf{s}_{[-k:]}\}}\frac{(r^j_{t})^{-1}}{\sum_{t'\in\mathcal{T}}(r^j_{t'})^{-1}}
\end{equation}
where $r^j_{t}$ is the rank of item $i_j$ produced by tool $t$.

\subsection{Inference and Contextual Reranking}~\label{sec:rerank}
We further incorporate a lightweight reranking module that refines the top-ranked items based on their textual features, enhancing the semantic alignment between recommendations and user intent:
\begin{equation}
    y_{\text{rank}} = \text{LLM}_{\text{rank}}(y_{\text{prof}},\text{desc}(f_\text{top}(\hat{\mathbf{r}},k'_{\text{rank}})),\text{desc}(\mathbf{s}_{[-k':]}))
\end{equation}
where $f_\text{top}(\hat{\mathbf{r}},k'_{\text{rank}})$ denotes the top $k'_{\text{rank}}$ items in the aggregated ranking list, and $\mathbf{s}_{[-k':]}$ denotes the most recent $k'$ items interacted with by the user. The output $y_{\text{rank}}$ is a reranked list of these $k'_{\text{rank}}$ items, which serves as the final ranked list. This reranking allows the agent to make fine-grained adjustments based on nuanced semantic signals that traditional ranking models often fail to capture.

\section{Experiment}

\begin{table*}[t]\caption{Overall performance comparison. The best and second-best methods are in boldface and underlined.}\label{tab:overall}
\
\resizebox{0.95\textwidth}{!}{
\begin{tabular}{l|rrrr|rrrr|rrrr}
\hline
\hline
Dataset    & \multicolumn{4}{c|}{Amazon}                                                                                         & \multicolumn{4}{c|}{Yelp}                                                                                            & \multicolumn{4}{c}{Goodreads}                                                                                           \\ \hline
Metric     & \multicolumn{1}{c}{R@10}     & \multicolumn{1}{c}{N@10}    & \multicolumn{1}{c}{R@20}     & \multicolumn{1}{c|}{N@20}    & \multicolumn{1}{c}{R@10}     & \multicolumn{1}{c}{N@10}    & \multicolumn{1}{c}{R@20}     & \multicolumn{1}{c|}{N@20}    & \multicolumn{1}{c}{R@10}     & \multicolumn{1}{c}{N@10}    & \multicolumn{1}{c}{R@20}      & \multicolumn{1}{c}{N@20}     \\ \hline
LightGCN   & 0.0250                      & 0.0111                      & 0.0375                     & 0.0143                       & 0.0438                      & 0.0216                     & 0.0750                      & 0.0293                       & 0.0625                      & 0.0244                      & 0.0875                       & 0.0309                       \\
SASRec      & 0.0438                      & 0.0206                      & 0.0813                      & 0.0301                       & 0.0688                      & 0.0413                      & 0.1063                      & 0.0508                       & \uline{0.1000}                      & \uline{0.0539}                     & \uline{0.1688}                      & \uline{0.0714}                       \\
SimpleX    & 0.0188                      & 0.0147                      & 0.0313                      & 0.0179                       & 0.0500                      & 0.0232                      & 0.0750                      & 0.0299                       & 0.0438                     & 0.0227                      & 0.0500                       & 0.0243                       \\
ENMF       & 0.0375                      & 0.0261                      & 0.0500                      & 0.0292                       & 0.0313                      & 0.0185                      & 0.0313                      & 0.0185                       & 0.0625                      & 0.0319                     & 0.0688                       & 0.0336                       \\
DiffRec & 0.0438                      & 0.0199                & 0.0438                & 0.0199                               & \uline{0.0813}                & \uline{0.0480}                     & \uline{0.1125}                & \uline{0.0587}                 & 0.0688                     & 0.0331                       & 0.1000                      & 0.0408                 \\
FEARec & 0.0500                &0.0189                     &\uline{0.1063}                     & \uline{0.0331}                      & 0.0688                     & 0.0305                     & 0.1063                     & 0.0339                       & 0.0563                      & 0.0283                      & 0.1188                       & 0.0439                       \\ \hline
BM25 & 0.0063                      & 0.0031                & 0.0063                & 0.0031                 & 0.0000                & 0.0000                      & 0.0000                & 0.0000                 & 0.0000                      & 0.0000                & 0.0063                 & 0.0014                 \\
LLMRank & 0.0125                & 0.0089                      & 0.0125                      & 0.0094                       & 0.0000                      & 0.0000                      & 0.0063                     & 0.0016                       & 0.0125                      & 0.0044                      & 0.0125                       & 0.0063                      \\ 
LightGCN$_{RAG}$ & 0.0250                      & 0.0141                      & 0.0438                      & 0.0137                       & 0.0375                      & 0.0185                      & 0.0750                      & 0.0347                       & 0.0188                      & 0.0120                      & 0.0250                       & 0.0118                       \\ 
SASRec$_{RAG}$ & \uline{0.0625}                     & \uline{0.0322}                      & 0.0750                      & 0.0266                       & 0.0625                      & 0.0282                      & 0.0750                      & 0.0377                       & 0.0188                      & 0.0108                      & 0.0188                      & 0.0113                        \\
SimpleX$_{RAG}$ & 0.0438                & 0.0202                      & 0.0563                      & 0.0247                           & 0.0500                   & 0.0284                      & 0.1000                      & 0.0437                       & 0.0063                      & 0.0063                      & 0.0188                       & 0.0077                       \\ \hline
\modelname      & \textbf{0.1000}       & \textbf{0.0535}             & \textbf{0.1125}             & \textbf{0.0542}              & \textbf{0.1063}             & \textbf{0.0598}             & \textbf{0.1250}             & \textbf{0.0614}              & \textbf{0.1688}             & \textbf{0.0809}             & \textbf{0.1813}             & \textbf{0.0834}              \\ \hline
Improv.    & \multicolumn{1}{l}{60.00\%} & \multicolumn{1}{l}{66.15\%} & \multicolumn{1}{l}{11.51\%} & \multicolumn{1}{l|}{14.28\%} & \multicolumn{1}{l}{30.75\%} & \multicolumn{1}{l}{24.58\%} & \multicolumn{1}{l}{11.11\%} & \multicolumn{1}{l|}{4.60\%} & \multicolumn{1}{l}{68.80\%} & \multicolumn{1}{l}{50.09\%} & \multicolumn{1}{l}{7.41\%} & \multicolumn{1}{l}{16.81\%} \\ \hline
\end{tabular}
}
\end{table*}

\subsection{Experiment Settings}
\subsubsection{Datasets}
We conduct experiments on three publicly available datasets: Amazon~\cite{dataset:amazon}, Yelp~\cite{dataset:yelp} and Goodreads~\cite{dataset:goodreads}.
Amazon dataset contains 1,054 users and 1,839 items under the category of movies and CDs with 35,492 interactions; Yelp contains 4,303 users and 3,429 businesses with 258,486 interactions; Goodreads contains 3,204 users and 2,167 books with 246,034 interactions. For each dataset, we chronologically organize users' interactions based on timestamps to construct their historical interaction sequences. Given that the framework is evaluated on a sequential recommendation task, we adopt the leave-one-out strategy for training as in prior works~\cite{sasrec,fearec}.

\subsubsection{Implementations and Cost Analysis}
We initialize the learning rates $\alpha$ and $\beta$ as $0.1$, and decay both by a factor of $0.5$ after each training epoch.
Since the evaluation metrics consider only the top 20 items, the numbers of recalled and reranked candidates, $k_{\text{eval}}$, $k_{\text{comp}}$, and $k'_{\text{rank}}$, are all set to 20.
Following prior LLM-agent approaches~\cite{agentcf,recagent,interecagent}, we train the recommendation tools with a fixed random seed on the \emph{full} training set covering all users. The randomly sampled 160 agents per dataset are used \emph{only} for the LLM-agent stage, i.e., optimizing and evaluating the per-user PT memory; they do not restrict the recommendation tools, which are trained on all users and rank over the entire item catalog for every user. Since each user's PT memory is optimized independently through the reflection modules, this stage is embarrassingly parallel and scales linearly with the number of users, so the sample size is a cost-control choice for LLM calls rather than an inherent scalability limit.
We adopt a quantized version of Phi-4~\cite{phi4} as the backbone LLM, deployed locally via vLLM with temperature set to 0.
The number of LLM API calls includes: (i) user profile generation, (ii) local evaluation of three tools together with global comparison during training, and (iii) contextual reranking during inference.
Accordingly, the total number of calls is $4nt + 2n$, where $n$ and $t$ denote the numbers of users and training epochs, respectively.
With the updated memory, \modelname requires less than 0.5 seconds per user to generate the final reranked list on a single V100-32GB GPU.
Although LLM-based agents generally incur higher inference latency than embedding-based recommendation models, they remain practical for offline ranking or as a second-stage reranking component in recommendation pipelines.

\subsubsection{Baselines}
We compare \modelname against each of its tools and other baselines from RecBole~\cite{recbole}, including ENMF~\cite{enmf}, DiffRec~\cite{diffrec}, and FEARec~\cite{fearec}. Besides, BM25~\cite{bm25} ranking items based on textual similarity and LLMRanker~\cite{llmrank} using the LLM as a zero-shot ranker are also compared as language-based baselines.
Following the idea of retrieval-augmented generation (RAG)~\cite{LewisPPPKGKLYR020}, we also compare \modelname with tool variants $\text{LightGCN}_{RAG}$, $\text{SASRec}_{RAG}$ and $\text{SimpleX}_{RAG}$. In these variants, LLMs are prompted to rerank the top-50 items from the ranked list of the corresponding tool. Since most prior LLM-agent works~\cite{agentcf,recmind,interecagent,agent4rec,recagent,iagent} are unable to rank all items in datasets, we exclude them from the comparison table instead of reporting trivially low full-ranking performance. Recall (R@10,20) and NDCG (N@10,20) are used as evaluation metrics.

\subsection{Overall Performance}
Performance comparison between \modelname and other baselines is shown in Table~\ref{tab:overall}.
First, \modelname consistently achieves the best performance across all datasets with notable improvements. This highlights the benefit of delegating full-ranking to specialized recommendation models while using the LLM for personalized tool selection and lightweight contextual reranking, which is particularly effective at top-ranked positions.
Second, language-based methods such as BM25 and LLMRank perform poorly in full-ranking settings, as they fail to model collaborative and sequential behavioral signals. Their reliance on textual similarity or general world knowledge limits their effectiveness in large item spaces.
Third, RAG baselines like LightGCN$_{RAG}$ do not consistently outperform their base models, since LLMs applied to candidates from a single retriever operate on potentially biased contexts. In contrast, \modelname improves robustness by reranking a small set of high-confidence candidates aggregated from multiple tools.

\subsection{Ablation Study}
\begin{figure}[]
    \begin{subfigure}{0.163\textwidth}
    \includegraphics[width=\textwidth]{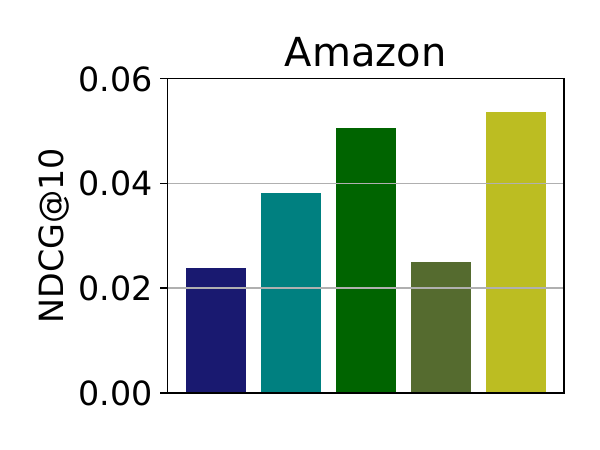}
    \end{subfigure}
    \hspace{-3mm}
    % \hfill
    \begin{subfigure}{0.163\textwidth}
    \includegraphics[width=\textwidth]{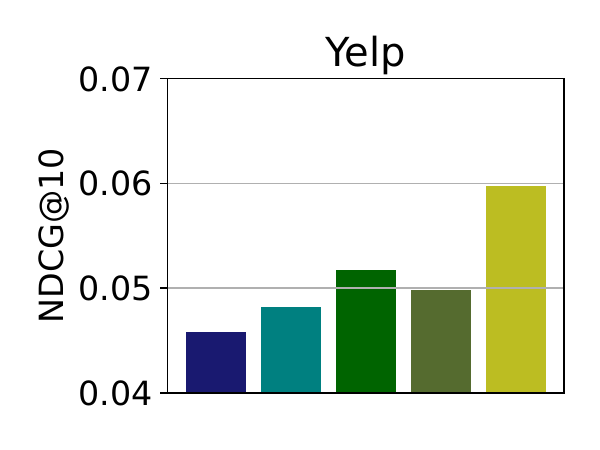}
    \end{subfigure}
    \hspace{-3mm}
    % \hfill
    \begin{subfigure}{0.163\textwidth}
    \includegraphics[width=\textwidth]{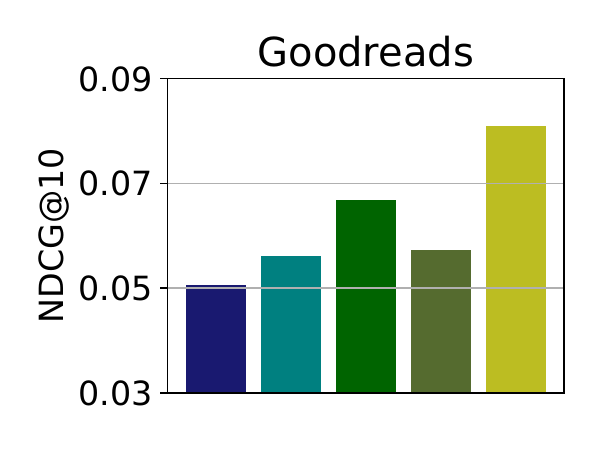}
    \end{subfigure}
    \begin{subfigure}{0.45\textwidth}
    \centering
    \vspace{-3mm}
    \includegraphics[width=\textwidth]{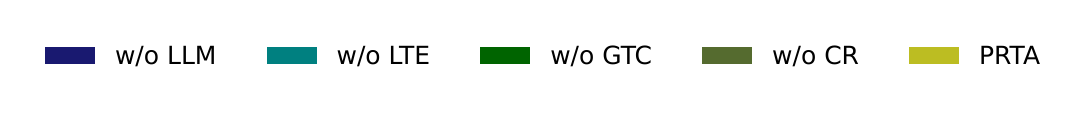}
    \end{subfigure}
    \vspace{-4mm}
        \caption{Performance of \modelname without each LLM-based
module. The bar of w/o LLM means the agent is only optimized
by ranking comparison without inference reranking.}~\label{fig:ablation}
    \vspace{-3mm}
\end{figure}
Figure~\ref{fig:ablation} shows the ablation results of \modelname by removing different LLM-based modules, including local tool evaluation (\textit{w/o LTE}), global tool comparison (\textit{w/o GTC}), and contextual reranking (\textit{w/o CR}). Removing all LLM modules (\textit{w/o LLM}) results in performance close to the single-tool setting in Table~\ref{tab:overall}. Contextual reranking provides substantial improvements on the Amazon dataset, where item semantics in user behavior sequences are more consistent and informative, enabling better personalization. In contrast, when textual information in behavior sequences is less indicative of user preference, as in Yelp, the benefit of reranking becomes smaller than that of local tool evaluation. Global tool comparison provides limited but stable improvements across datasets, indicating partial functional overlap with ranking-based selection while still contributing complementary signals.

\subsection{PT Memory Visualization}
\begin{figure}[]
    \begin{subfigure}{0.163\textwidth}
    \includegraphics[width=\textwidth]{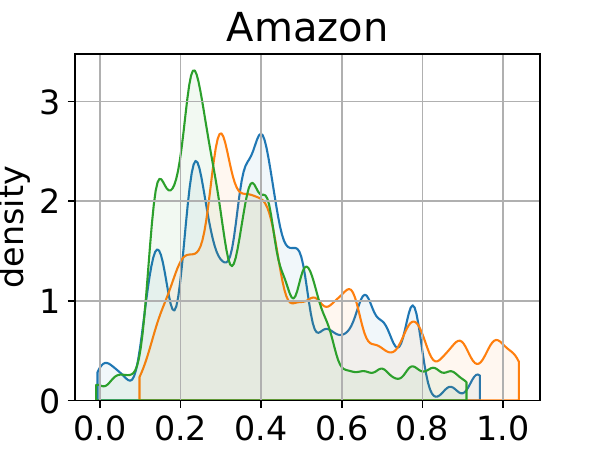}
    \end{subfigure}
    \hspace{-3mm}
    % \hfill
    \begin{subfigure}{0.163\textwidth}
    \includegraphics[width=\textwidth]{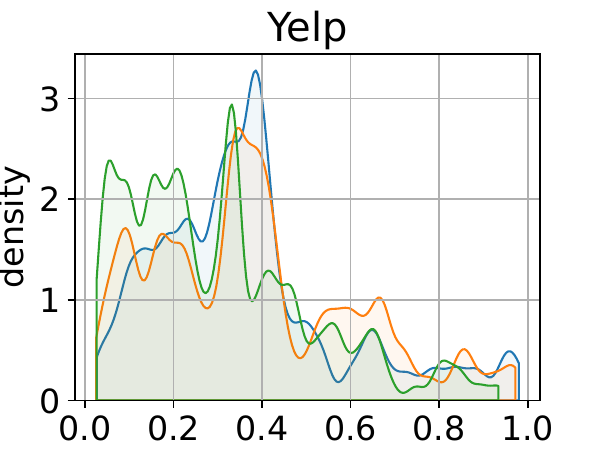}
    \end{subfigure}
    \hspace{-3mm}
    % \hfill
    \begin{subfigure}{0.163\textwidth}
    \includegraphics[width=\textwidth]{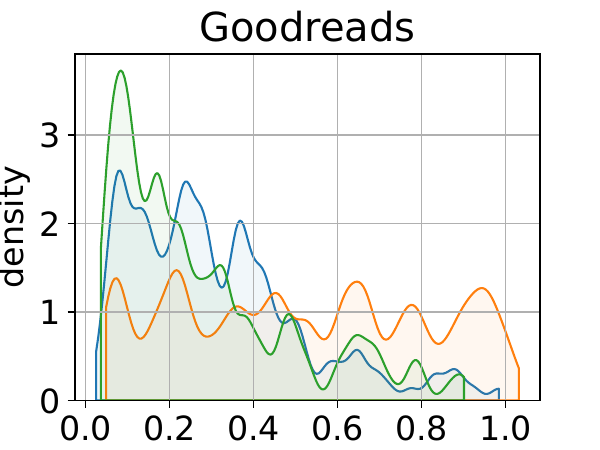}
    \end{subfigure}
    \begin{subfigure}{0.35\textwidth}
    \centering
    \vspace{-3mm}
    \includegraphics[width=\textwidth]{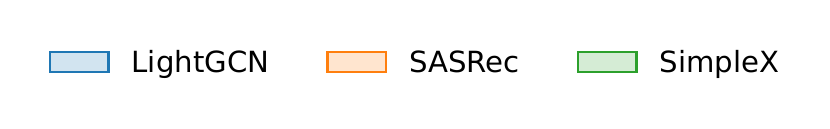}
    \end{subfigure}
    \vspace{-4mm}
        \caption{The KDE plot visualizing the distributions of the tool weights in PT memory. Each curve denotes the distributions of the weights of the corresponding tool for all users.}
  \label{fig:kde}
\end{figure}

To illustrate the distribution of the optimized tool weights stored in PT memory, we plot KDE curves for all users on three datasets in Figure~\ref{fig:kde}. The x-axis represents tool weight values and the y-axis denotes their density, where higher density indicates that more user agents assign the corresponding weight to a given tool. We observe that the weights of SASRec are more concentrated near 1 compared to LightGCN and SimpleX, suggesting that sequence-based recommendations are more consistent with users' historical behaviors. In contrast, the weights of SimpleX are more concentrated toward 0, indicating that MF-based recommendations are less semantically consistent with user preferences and therefore receive lower weights from the LLM agent.
This trend is particularly evident on Goodreads, where many books are consumed in series. 

% \begin{figure}[]
%     \begin{subfigure}{0.163\textwidth}
%     \includegraphics[width=\textwidth]{figures/tool_weights/piechart_amazon.pdf}
%     \end{subfigure}
%     \hspace{-3mm}
%     % \hfill
%     \begin{subfigure}{0.163\textwidth}
%     \includegraphics[width=\textwidth]{figures/tool_weights/piechart_yelp.pdf}
%     \end{subfigure}
%     \hspace{-3mm}
%     % \hfill
%     \begin{subfigure}{0.163\textwidth}
%     \includegraphics[width=\textwidth]{figures/tool_weights/piechart_goodreads.pdf}
%     \end{subfigure}
%     \begin{subfigure}{0.35\textwidth}
%     \centering
%     \vspace{-3mm}
%     \includegraphics[width=\textwidth]{figures/tool_weights/tool_distribution_legend.pdf}
%     \end{subfigure}
%     \vspace{-4mm}
%         \caption{Ratio of users w.r.t their dominative tools. The dominative tool for a user is the tool with the highest weight out of the three tools after training.}
%   \label{fig:dominate_tool}
% \end{figure}

\section{Conclusion}\label{sec:conclusion}
In this work, we propose a modular framework, \modelname, for full-ranking recommendation by decoupling behavioral modeling from semantic reasoning. By orchestrating multiple specialized recommendation models as tools via a centralized LLM-based agent, \modelname combines the strengths of traditional recommenders and language-based reasoning, circumventing the hallucination and context-length limitations of LLMs by architectural design rather than resolving them within the model itself.
The proposed reflection modules enable personalized tool selection, while lightweight contextual reranking further improves recommendation quality. 

\begin{acks}

This work is supported by NSFC through grants U25B2029 and 62322202, Beijing Natural Science Foundation through grant L253021, the Pioneer and Leading Goose R\&D Program of Zhejiang through grant 2025C02044,  
S\&T Program of Hebei through grant 26280103D, 
Science Research Project of Hebei Higher Education Institutions under grant CYZD2026005, and Major Science and Technology Special Projects of Yunnan Province through grants 202502AD080012 and 202502AD080006.

\end{acks}
\bibliographystyle{ACM-Reference-Format}
\balance
\bibliography{LRTA}

@inproceedings{WuZHHYCISLZ24,
  author       = {Shirley Wu and
                  Shiyu Zhao and
                  Qian Huang and
                  Kexin Huang and
                  Michihiro Yasunaga and
                  Kaidi Cao and
                  Vassilis N. Ioannidis and
                  Karthik Subbian and
                  Jure Leskovec and
                  James Y. Zou},
  editor       = {Amir Globersons and
                  Lester Mackey and
                  Danielle Belgrave and
                  Angela Fan and
                  Ulrich Paquet and
                  Jakub M. Tomczak and
                  Cheng Zhang},
  title        = {AvaTaR: Optimizing {LLM} Agents for Tool Usage via Contrastive Reasoning},
  booktitle    = {Advances in Neural Information Processing Systems 38: Annual Conference
                  on Neural Information Processing Systems 2024, NeurIPS 2024, Vancouver,
                  BC, Canada, December 10 - 15, 2024},
  year         = {2024},
  url          = {http://papers.nips.cc/paper\_files/paper/2024/hash/2db8ce969b000fe0b3fb172490c33ce8-Abstract-Conference.html},
  bibsource    = {dblp computer science bibliography, https://dblp.org}
}

@inproceedings{simplex,
  author       = {Kelong Mao and
                  Jieming Zhu and
                  Jinpeng Wang and
                  Quanyu Dai and
                  Zhenhua Dong and
                  Xi Xiao and
                  Xiuqiang He},
  editor       = {Gianluca Demartini and
                  Guido Zuccon and
                  J. Shane Culpepper and
                  Zi Huang and
                  Hanghang Tong},
  title        = {SimpleX: {A} Simple and Strong Baseline for Collaborative Filtering},
  booktitle    = {{CIKM} '21: The 30th {ACM} International Conference on Information
                  and Knowledge Management, Virtual Event, Queensland, Australia, November
                  1 - 5, 2021},
  pages        = {1243--1252},
  publisher    = {{ACM}},
  year         = {2021},
  url          = {https://doi.org/10.1145/3459637.3482297},
  doi          = {10.1145/3459637.3482297},
  bibsource    = {dblp computer science bibliography, https://dblp.org}
}

@inproceedings{sasrec,
  author       = {Wang{-}Cheng Kang and
                  Julian J. McAuley},
  title        = {Self-Attentive Sequential Recommendation},
  booktitle    = {{IEEE} International Conference on Data Mining, {ICDM} 2018, Singapore,
                  November 17-20, 2018},
  pages        = {197--206},
  publisher    = {{IEEE} Computer Society},
  year         = {2018},
  url          = {https://doi.org/10.1109/ICDM.2018.00035},
  doi          = {10.1109/ICDM.2018.00035},
  bibsource    = {dblp computer science bibliography, https://dblp.org}
}

@inproceedings{HuaGXJLZ24,
  author       = {Wenyue Hua and
                  Yingqiang Ge and
                  Shuyuan Xu and
                  Jianchao Ji and
                  Zelong Li and
                  Yongfeng Zhang},
  editor       = {Yvette Graham and
                  Matthew Purver},
  title        = {{UP5:} Unbiased Foundation Model for Fairness-aware Recommendation},
  booktitle    = {Proceedings of the 18th Conference of the European Chapter of the
                  Association for Computational Linguistics, {EACL} 2024 - Volume 1:
                  Long Papers, St. Julian's, Malta, March 17-22, 2024},
  pages        = {1899--1912},
  publisher    = {Association for Computational Linguistics},
  year         = {2024},
  url          = {https://aclanthology.org/2024.eacl-long.114},
  bibsource    = {dblp computer science bibliography, https://dblp.org}
}

@inproceedings{toolrec,
  author       = {Yuyue Zhao and
                  Jiancan Wu and
                  Xiang Wang and
                  Wei Tang and
                  Dingxian Wang and
                  Maarten de Rijke},
  editor       = {Grace Hui Yang and
                  Hongning Wang and
                  Sam Han and
                  Claudia Hauff and
                  Guido Zuccon and
                  Yi Zhang},
  title        = {Let Me Do It For You: Towards {LLM} Empowered Recommendation via Tool
                  Learning},
  booktitle    = {Proceedings of the 47th International {ACM} {SIGIR} Conference on
                  Research and Development in Information Retrieval, {SIGIR} 2024, Washington
                  DC, USA, July 14-18, 2024},
  pages        = {1796--1806},
  publisher    = {{ACM}},
  year         = {2024},
  url          = {https://doi.org/10.1145/3626772.3657828},
  doi          = {10.1145/3626772.3657828},
  bibsource    = {dblp computer science bibliography, https://dblp.org}
}

@inproceedings{recmind,
    title = "{R}ec{M}ind: Large Language Model Powered Agent For Recommendation",
    author = "Wang, Yancheng  and
      Jiang, Ziyan  and
      Chen, Zheng  and
      Yang, Fan  and
      Zhou, Yingxue  and
      Cho, Eunah  and
      Fan, Xing  and
      Lu, Yanbin  and
      Huang, Xiaojiang  and
      Yang, Yingzhen",
    editor = "Duh, Kevin  and
      Gomez, Helena  and
      Bethard, Steven",
    booktitle = "Findings of the Association for Computational Linguistics: NAACL 2024",
    month = jun,
    year = "2024",
    address = "Mexico City, Mexico",
    publisher = "Association for Computational Linguistics",
    url = "https://aclanthology.org/2024.findings-naacl.271/",
    doi = "10.18653/v1/2024.findings-naacl.271",
    pages = "4351--4364",
}

@article{recagent,
author = {Wang, Lei and Zhang, Jingsen and Yang, Hao and Chen, Zhi-Yuan and Tang, Jiakai and Zhang, Zeyu and Chen, Xu and Lin, Yankai and Sun, Hao and Song, Ruihua and Zhao, Xin and Xu, Jun and Dou, Zhicheng and Wang, Jun and Wen, Ji-Rong},
title = {User Behavior Simulation with Large Language Model-based Agents},
year = {2025},
issue_date = {March 2025},
publisher = {Association for Computing Machinery},
address = {New York, NY, USA},
volume = {43},
number = {2},
issn = {1046-8188},
url = {https://doi.org/10.1145/3708985},
doi = {10.1145/3708985},
journal = {ACM Trans. Inf. Syst.},
month = jan,
articleno = {55},
numpages = {37}
}

@inproceedings{tallrec,
  author       = {Keqin Bao and
                  Jizhi Zhang and
                  Yang Zhang and
                  Wenjie Wang and
                  Fuli Feng and
                  Xiangnan He},
  editor       = {Jie Zhang and
                  Li Chen and
                  Shlomo Berkovsky and
                  Min Zhang and
                  Tommaso Di Noia and
                  Justin Basilico and
                  Luiz Pizzato and
                  Yang Song},
  title        = {TALLRec: An Effective and Efficient Tuning Framework to Align Large
                  Language Model with Recommendation},
  booktitle    = {Proceedings of the 17th {ACM} Conference on Recommender Systems, RecSys
                  2023, Singapore, Singapore, September 18-22, 2023},
  pages        = {1007--1014},
  publisher    = {{ACM}},
  year         = {2023},
  url          = {https://doi.org/10.1145/3604915.3608857},
  doi          = {10.1145/3604915.3608857},
  bibsource    = {dblp computer science bibliography, https://dblp.org}
}

@inproceedings{agentcf,
  author       = {Junjie Zhang and
                  Yupeng Hou and
                  Ruobing Xie and
                  Wenqi Sun and
                  Julian J. McAuley and
                  Wayne Xin Zhao and
                  Leyu Lin and
                  Ji{-}Rong Wen},
  editor       = {Tat{-}Seng Chua and
                  Chong{-}Wah Ngo and
                  Ravi Kumar and
                  Hady W. Lauw and
                  Roy Ka{-}Wei Lee},
  title        = {AgentCF: Collaborative Learning with Autonomous Language Agents for
                  Recommender Systems},
  booktitle    = {Proceedings of the {ACM} on Web Conference 2024, {WWW} 2024, Singapore,
                  May 13-17, 2024},
  pages        = {3679--3689},
  publisher    = {{ACM}},
  year         = {2024},
  url          = {https://doi.org/10.1145/3589334.3645537},
  doi          = {10.1145/3589334.3645537},
  bibsource    = {dblp computer science bibliography, https://dblp.org}
}

@inproceedings{agent4rec,
  author       = {An Zhang and
                  Yuxin Chen and
                  Leheng Sheng and
                  Xiang Wang and
                  Tat{-}Seng Chua},
  editor       = {Grace Hui Yang and
                  Hongning Wang and
                  Sam Han and
                  Claudia Hauff and
                  Guido Zuccon and
                  Yi Zhang},
  title        = {On Generative Agents in Recommendation},
  booktitle    = {Proceedings of the 47th International {ACM} {SIGIR} Conference on
                  Research and Development in Information Retrieval, {SIGIR} 2024, Washington
                  DC, USA, July 14-18, 2024},
  pages        = {1807--1817},
  publisher    = {{ACM}},
  year         = {2024},
  url          = {https://doi.org/10.1145/3626772.3657844},
  doi          = {10.1145/3626772.3657844},
  bibsource    = {dblp computer science bibliography, https://dblp.org}
}

@inproceedings{Toolformer,
  author       = {Timo Schick and
                  Jane Dwivedi{-}Yu and
                  Roberto Dess{\`{\i}} and
                  Roberta Raileanu and
                  Maria Lomeli and
                  Eric Hambro and
                  Luke Zettlemoyer and
                  Nicola Cancedda and
                  Thomas Scialom},
  editor       = {Alice Oh and
                  Tristan Naumann and
                  Amir Globerson and
                  Kate Saenko and
                  Moritz Hardt and
                  Sergey Levine},
  title        = {Toolformer: Language Models Can Teach Themselves to Use Tools},
  booktitle    = {Advances in Neural Information Processing Systems 36: Annual Conference
                  on Neural Information Processing Systems 2023, NeurIPS 2023, New Orleans,
                  LA, USA, December 10 - 16, 2023},
  year         = {2023},
  url          = {http://papers.nips.cc/paper\_files/paper/2023/hash/d842425e4bf79ba039352da0f658a906-Abstract-Conference.html},
  bibsource    = {dblp computer science bibliography, https://dblp.org}
}

@inproceedings{HuggingGPT,
  author       = {Yongliang Shen and
                  Kaitao Song and
                  Xu Tan and
                  Dongsheng Li and
                  Weiming Lu and
                  Yueting Zhuang},
  editor       = {Alice Oh and
                  Tristan Naumann and
                  Amir Globerson and
                  Kate Saenko and
                  Moritz Hardt and
                  Sergey Levine},
  title        = {HuggingGPT: Solving {AI} Tasks with ChatGPT and its Friends in Hugging
                  Face},
  booktitle    = {Advances in Neural Information Processing Systems 36: Annual Conference
                  on Neural Information Processing Systems 2023, NeurIPS 2023, New Orleans,
                  LA, USA, December 10 - 16, 2023},
  year         = {2023},
  url          = {http://papers.nips.cc/paper\_files/paper/2023/hash/77c33e6a367922d003ff102ffb92b658-Abstract-Conference.html},
  bibsource    = {dblp computer science bibliography, https://dblp.org}
}

@inproceedings{pcl,
author = {Yang, Mingdai and Yang, Fan and Guo, Yanhui and Xu, Shaoyuan and Zhou, Tianchen and Chen, Yetian and Shao, Simone and Liu, Jia and Gao, Yan},
title = {PCL: Prompt-based Continual Learning for User Modeling in Recommender Systems},
year = {2025},
isbn = {9798400713316},
publisher = {Association for Computing Machinery},
address = {New York, NY, USA},
url = {https://doi.org/10.1145/3701716.3715589},
doi = {10.1145/3701716.3715589},
booktitle = {Companion Proceedings of the ACM on Web Conference 2025},
pages = {1475–1479},
numpages = {5},
location = {Sydney NSW, Australia},
series = {WWW '25}
}

@inproceedings{ReAct,
  author       = {Shunyu Yao and
                  Jeffrey Zhao and
                  Dian Yu and
                  Nan Du and
                  Izhak Shafran and
                  Karthik R. Narasimhan and
                  Yuan Cao},
  title        = {ReAct: Synergizing Reasoning and Acting in Language Models},
  booktitle    = {The Eleventh International Conference on Learning Representations,
                  {ICLR} 2023, Kigali, Rwanda, May 1-5, 2023},
  publisher    = {OpenReview.net},
  year         = {2023},
  url          = {https://openreview.net/forum?id=WE\_vluYUL-X},
  bibsource    = {dblp computer science bibliography, https://dblp.org}
}

@inproceedings{agentdr,
author = {Yang, Mingdai and Choudhary, Nurendra and Du, Jiangshu and Huang, Edward W and Yu, Philip and Subbian, Karthik and Koutra, Danai},
title = {AgentDR: Dynamic Recommendation with Implicit Item-Item Relations via LLM-based Agents},
year = {2026},
isbn = {9798400723070},
publisher = {Association for Computing Machinery},
address = {New York, NY, USA},
url = {https://doi.org/10.1145/3774904.3792304},
doi = {10.1145/3774904.3792304},
booktitle = {Proceedings of the ACM Web Conference 2026},
pages = {6230–6241},
numpages = {12},
location = {United Arab Emirates},
series = {WWW '26}
}

@article{phi4,
  author       = {Marah I Abdin and
                  Jyoti Aneja and
                  Harkirat S. Behl and
                  S{\'{e}}bastien Bubeck and
                  Ronen Eldan and
                  Suriya Gunasekar and
                  Michael Harrison and
                  Russell J. Hewett and
                  Mojan Javaheripi and
                  Piero Kauffmann and
                  James R. Lee and
                  Yin Tat Lee and
                  Yuanzhi Li and
                  Weishung Liu and
                  Caio C. T. Mendes and
                  Anh Nguyen and
                  Eric Price and
                  Gustavo de Rosa and
                  Olli Saarikivi and
                  Adil Salim and
                  Shital Shah and
                  Xin Wang and
                  Rachel Ward and
                  Yue Wu and
                  Dingli Yu and
                  Cyril Zhang and
                  Yi Zhang},
  title        = {Phi-4 Technical Report},
  journal      = {CoRR},
  volume       = {abs/2412.08905},
  year         = {2024},
  url          = {https://doi.org/10.48550/arXiv.2412.08905},
  doi          = {10.48550/ARXIV.2412.08905},
  eprinttype   = {arXiv},
  eprint       = {2412.08905},
  bibsource    = {dblp computer science bibliography, https://dblp.org}
}

@inproceedings{recbole,
author = {Zhao, Wayne Xin and Mu, Shanlei and Hou, Yupeng and Lin, Zihan and Chen, Yushuo and Pan, Xingyu and Li, Kaiyuan and Lu, Yujie and Wang, Hui and Tian, Changxin and Min, Yingqian and Feng, Zhichao and Fan, Xinyan and Chen, Xu and Wang, Pengfei and Ji, Wendi and Li, Yaliang and Wang, Xiaoling and Wen, Ji-Rong},
title = {RecBole: Towards a Unified, Comprehensive and Efficient Framework for Recommendation Algorithms},
year = {2021},
isbn = {9781450384469},
publisher = {Association for Computing Machinery},
address = {New York, NY, USA},
url = {https://doi.org/10.1145/3459637.3482016},
doi = {10.1145/3459637.3482016},
booktitle = {Proceedings of the 30th ACM International Conference on Information \& Knowledge Management},
pages = {4653–4664},
numpages = {12},
location = {Virtual Event, Queensland, Australia},
series = {CIKM '21}
}

@misc{dataset:yelp,
  title        = {Yelp Open Dataset},
  author       = {{Yelp}},
  year         = {2015},
  howpublished = {\url{https://business.yelp.com/data/resources/open-dataset/}},
  note         = {Accessed: 2025-02}
}

@inproceedings{iagent,
    title = "i{A}gent: {LLM} Agent as a Shield between User and Recommender Systems",
    author = "Xu, Wujiang  and
      Shi, Yunxiao  and
      Liang, Zujie  and
      Ning, Xuying  and
      Mei, Kai  and
      Wang, Kun  and
      Zhu, Xi  and
      Xu, Min  and
      Zhang, Yongfeng",
   editor = "Che, Wanxiang  and
      Nabende, Joyce  and
      Shutova, Ekaterina  and
      Pilehvar, Mohammad Taher",
    booktitle = "Findings of the Association for Computational Linguistics: ACL 2025",
    month = jul,
    year = "2025",
    address = "Vienna, Austria",
    publisher = "Association for Computational Linguistics",
    pages = "18056--18084",
    ISBN = "979-8-89176-256-5"
}

@article{JiLFYSXIBMF23,
  author       = {Ziwei Ji and
                  Nayeon Lee and
                  Rita Frieske and
                  Tiezheng Yu and
                  Dan Su and
                  Yan Xu and
                  Etsuko Ishii and
                  Yejin Bang and
                  Andrea Madotto and
                  Pascale Fung},
  title        = {Survey of Hallucination in Natural Language Generation},
  journal      = {{ACM} Comput. Surv.},
  volume       = {55},
  number       = {12},
  pages        = {248:1--248:38},
  year         = {2023},
  url          = {https://doi.org/10.1145/3571730},
  doi          = {10.1145/3571730},
  bibsource    = {dblp computer science bibliography, https://dblp.org}
}

@inproceedings{JiangWMCWWZ25,
  author       = {Chumeng Jiang and
                  Jiayin Wang and
                  Weizhi Ma and
                  Charles L. A. Clarke and
                  Shuai Wang and
                  Chuhan Wu and
                  Min Zhang},
  editor       = {Guodong Long and
                  Michale Blumestein and
                  Yi Chang and
                  Liane Lewin{-}Eytan and
                  Zi Helen Huang and
                  Elad Yom{-}Tov},
  title        = {Beyond Utility: Evaluating {LLM} as Recommender},
  booktitle    = {Proceedings of the {ACM} on Web Conference 2025, {WWW} 2025, Sydney,
                  NSW, Australia, 28 April 2025- 2 May 2025},
  pages        = {3850--3862},
  publisher    = {{ACM}},
  year         = {2025},
  url          = {https://doi.org/10.1145/3696410.3714759},
  doi          = {10.1145/3696410.3714759},
  bibsource    = {dblp computer science bibliography, https://dblp.org}
}

@inproceedings{llmrank,
  author       = {Yupeng Hou and
                  Junjie Zhang and
                  Zihan Lin and
                  Hongyu Lu and
                  Ruobing Xie and
                  Julian J. McAuley and
                  Wayne Xin Zhao},
  editor       = {Nazli Goharian and
                  Nicola Tonellotto and
                  Yulan He and
                  Aldo Lipani and
                  Graham McDonald and
                  Craig Macdonald and
                  Iadh Ounis},
  title        = {Large Language Models are Zero-Shot Rankers for Recommender Systems},
  booktitle    = {Advances in Information Retrieval - 46th European Conference on Information
                  Retrieval, {ECIR} 2024, Glasgow, UK, March 24-28, 2024, Proceedings,
                  Part {II}},
  series       = {Lecture Notes in Computer Science},
  volume       = {14609},
  pages        = {364--381},
  publisher    = {Springer},
  year         = {2024},
  url          = {https://doi.org/10.1007/978-3-031-56060-6\_24},
  doi          = {10.1007/978-3-031-56060-6\_24},
  bibsource    = {dblp computer science bibliography, https://dblp.org}
}

@inproceedings{dataset:amazon,
  author       = {Jianmo Ni and
                  Jiacheng Li and
                  Julian J. McAuley},
   title        = {Justifying Recommendations using Distantly-Labeled Reviews and Fine-Grained
                  Aspects},
  booktitle    = {Proceedings of the 2019 Conference on Empirical Methods in Natural
                  Language Processing and the 9th International Joint Conference on
                  Natural Language Processing},
  pages        = {188--197},
  publisher    = {Association for Computational Linguistics},
  year         = {2019},
}

@inproceedings{dataset:goodreads,
  author       = {Mengting Wan and
                  Rishabh Misra and
                  Ndapa Nakashole and
                  Julian J. McAuley},
  title        = {Fine-Grained Spoiler Detection from Large-Scale Review Corpora},
  booktitle    = {Proceedings of the 57th Conference of the Association for Computational
                  Linguistics},
  pages        = {2605--2610},
  publisher    = {Association for Computational Linguistics},
  year         = {2019},
}

@inproceedings{p5,
  author       = {Shijie Geng and
                  Shuchang Liu and
                  Zuohui Fu and
                  Yingqiang Ge and
                  Yongfeng Zhang},
   title        = {Recommendation as Language Processing {(RLP):} {A} Unified Pretrain,
                  Personalized Prompt {\&} Predict Paradigm {(P5)}},
  booktitle    = {RecSys '22: Sixteenth {ACM} Conference on Recommender Systems},
  pages        = {299--315},
  publisher    = {{ACM}},
  year         = {2022},
}

@inproceedings{ihp,
  author       = {Mingdai Yang and
                  Zhiwei Liu and
                  Liangwei Yang and
                  Xiaolong Liu and
                  Chen Wang and
                  Hao Peng and
                  Philip S. Yu},
  title        = {Instruction-based Hypergraph Pretraining},
  booktitle    = {Proceedings of the 47th International {ACM} {SIGIR} Conference on
                  Research and Development in Information Retrieval},
  pages        = {501--511},
  publisher    = {{ACM}},
  year         = {2024},
  bibsource    = {dblp computer science bibliography, https://dblp.org}
}

@inproceedings{LewisPPPKGKLYR020,
  author       = {Patrick Lewis and
                  Ethan Perez and
                  Aleksandra Piktus and
                  Fabio Petroni and
                  Vladimir Karpukhin and
                  Naman Goyal and
                  Heinrich K{\"{u}}ttler and
                  Mike Lewis and
                  Wen{-}tau Yih and
                  Tim Rockt{\"{a}}schel and
                  Sebastian Riedel and
                  Douwe Kiela},
  editor       = {Hugo Larochelle and
                  Marc'Aurelio Ranzato and
                  Raia Hadsell and
                  Maria{-}Florina Balcan and
                  Hsuan{-}Tien Lin},
  title        = {Retrieval-Augmented Generation for Knowledge-Intensive {NLP} Tasks},
  booktitle    = {Advances in Neural Information Processing Systems 33: Annual Conference
                  on Neural Information Processing Systems 2020, NeurIPS 2020, December
                  6-12, 2020, virtual},
  year         = {2020},
  url          = {https://proceedings.neurips.cc/paper/2020/hash/6b493230205f780e1bc26945df7481e5-Abstract.html},
  bibsource    = {dblp computer science bibliography, https://dblp.org}
}

@article{bm25,
  author       = {Stephen E. Robertson and
                  Hugo Zaragoza},
  title        = {The Probabilistic Relevance Framework: {BM25} and Beyond},
  journal      = {Found. Trends Inf. Retr.},
  volume       = {3},
  number       = {4},
  pages        = {333--389},
  year         = {2009},
  url          = {https://doi.org/10.1561/1500000019},
  doi          = {10.1561/1500000019},
  bibsource    = {dblp computer science bibliography, https://dblp.org}
}

@article{enmf,
  author       = {Chong Chen and
                  Min Zhang and
                  Yongfeng Zhang and
                  Yiqun Liu and
                  Shaoping Ma},
  title        = {Efficient Neural Matrix Factorization without Sampling for Recommendation},
  journal      = {{ACM} Trans. Inf. Syst.},
  volume       = {38},
  number       = {2},
  pages        = {14:1--14:28},
  year         = {2020},
  url          = {https://doi.org/10.1145/3373807},
  doi          = {10.1145/3373807},
  bibsource    = {dblp computer science bibliography, https://dblp.org}
}

@inproceedings{diffrec,
  author       = {Wenjie Wang and
                  Yiyan Xu and
                  Fuli Feng and
                  Xinyu Lin and
                  Xiangnan He and
                  Tat{-}Seng Chua},
  editor       = {Hsin{-}Hsi Chen and
                  Wei{-}Jou (Edward) Duh and
                  Hen{-}Hsen Huang and
                  Makoto P. Kato and
                  Josiane Mothe and
                  Barbara Poblete},
  title        = {Diffusion Recommender Model},
  booktitle    = {Proceedings of the 46th International {ACM} {SIGIR} Conference on
                  Research and Development in Information Retrieval, {SIGIR} 2023, Taipei,
                  Taiwan, July 23-27, 2023},
  pages        = {832--841},
  publisher    = {{ACM}},
  year         = {2023},
  url          = {https://doi.org/10.1145/3539618.3591663},
  doi          = {10.1145/3539618.3591663},
  bibsource    = {dblp computer science bibliography, https://dblp.org}
}

@inproceedings{fearec,
  author       = {Xinyu Du and
                  Huanhuan Yuan and
                  Pengpeng Zhao and
                  Jianfeng Qu and
                  Fuzhen Zhuang and
                  Guanfeng Liu and
                  Yanchi Liu and
                  Victor S. Sheng},
  editor       = {Hsin{-}Hsi Chen and
                  Wei{-}Jou (Edward) Duh and
                  Hen{-}Hsen Huang and
                  Makoto P. Kato and
                  Josiane Mothe and
                  Barbara Poblete},
  title        = {Frequency Enhanced Hybrid Attention Network for Sequential Recommendation},
  booktitle    = {Proceedings of the 46th International {ACM} {SIGIR} Conference on
                  Research and Development in Information Retrieval, {SIGIR} 2023, Taipei,
                  Taiwan, July 23-27, 2023},
  pages        = {78--88},
  publisher    = {{ACM}},
  year         = {2023},
  url          = {https://doi.org/10.1145/3539618.3591689},
  doi          = {10.1145/3539618.3591689},
  bibsource    = {dblp computer science bibliography, https://dblp.org}
}

@inproceedings{LightGCN,
  author       = {Xiangnan He and
                  Kuan Deng and
                  Xiang Wang and
                  Yan Li and
                  Yong{-}Dong Zhang and
                  Meng Wang},
  title        = {LightGCN: Simplifying and Powering Graph Convolution Network for Recommendation},
  booktitle    = {Proceedings of the 43rd International {ACM} {SIGIR} conference on
                  research and development in Information Retrieval},
  pages        = {639--648},
  publisher    = {{ACM}},
  year         = {2020},
}

@article{interecagent,
author = {Huang, Xu and Lian, Jianxun and Lei, Yuxuan and Yao, Jing and Lian, Defu and Xie, Xing},
title = {Recommender AI Agent: Integrating Large Language Models for Interactive Recommendations},
year = {2025},
publisher = {Association for Computing Machinery},
address = {New York, NY, USA},
issn = {1046-8188},
url = {https://doi.org/10.1145/3731446},
doi = {10.1145/3731446},
journal = {ACM Trans. Inf. Syst.},
month = apr
}

@inproceedings{0007XCW25,
  author       = {Hang Li and
                  Tianlong Xu and
                  Ethan Chang and
                  Qingsong Wen},
  editor       = {Toby Walsh and
                  Julie Shah and
                  Zico Kolter},
  title        = {Knowledge Tagging with Large Language Model Based Multi-Agent System},
  booktitle    = {AAAI-25, Sponsored by the Association for the Advancement of Artificial
                  Intelligence, February 25 - March 4, 2025, Philadelphia, PA, {USA}},
  pages        = {28775--28782},
  publisher    = {{AAAI} Press},
  year         = {2025},
  url          = {https://doi.org/10.1609/aaai.v39i28.35141},
  doi          = {10.1609/AAAI.V39I28.35141},
  bibsource    = {dblp computer science bibliography, https://dblp.org}
}

\end{document}